# Modeling Intercalated Group-4-Metal Nitride Halide Superconductivity with Interlayer Coulomb Coupling

**Dale R. Harshman** [1,2] · **Anthony T. Fiory** [3]

**Abstract**  Behavior consistent with Coulomb-mediated high-$T_C$ superconductivity is shown to be present in the intercalated group-4-metal nitride halides $A_x(S)_yMNX$, where the $MNX$ host ($M$ = Ti, Zr, Hf; $X$ = Cl, Br) is partially intercalated with cations $A_x$ and optionally molecular species $(S)_y$ in the van der Waals gap between the halide $X$ layers, expanding the basal-plane spacing $d$. The optimal transition temperature is modeled by $T_{C0} \propto \zeta^{-1}(\sigma/A)^{1/2}$, where the participating fractional charge per area per formula unit $\sigma/A$ and the distance $\zeta$, given by the transverse $A_x$-$X$ separation ($\zeta < d$), govern the interlayer Coulomb coupling. From experiment results for β-form compounds based on Zr and Hf, in which concentrations x of $A_x$ are varied, it is shown that $\sigma = \gamma[v(x_{opt} - x_0)]$, where $x_{opt}$ is the optimal doping, $x_0$ is the onset of superconducting behavior, v is the $A_x$ charge state, and $\gamma = 1/8$ is a factor determined by the model. Observations of $T_C < T_{C0}$ in the comparatively more disordered α-$A_x(S)_yTiNX$ compounds are modeled as pair-breaking by remote Coulomb scattering from the $A_x$ cations, which attenuates exponentially with increasing $\zeta$. The $T_{C0}$ values calculated for nine $A_x(S)_yMNCl$ compounds, shown to be optimal, agree with the measured $T_C$ to within experimental error. The model for $T_{C0}$ is also found to be consistent with the absence of high-$T_C$ characteristics for $A_xMNX$ compounds in which a spatially separated intercalation layer is not formed.

Dale R. Harshman
drh@physikon.net

Anthony T. Fiory
fiory@alum.mit.edu

[1] Physikon Research Corporation, Lynden, WA 98264, USA

[2] Department of Physics, The College of William and Mary, Williamsburg, VA 23187, USA

[3] Department of Physics, New Jersey Institute of Technology, Newark, NJ 07102, USA

# 1 Introduction

The present study considers the exotic superconductivity developed in $A_x(S)_yMNX$ compounds ($M$ denotes Ti, Zr, or Hf and $X$ denotes Cl or Br) comprising cationic elements $A_x$ and optionally molecular species $(S)_y$ intercalated between group-4 metal nitride halides $[MNX]_2$ layers, where $x \leq 1/2$ and $y \leq 1/2$. Since their discovery [1], numerous experimental studies have been undertaken in an effort to understand the somewhat complex behavior of intercalated $MNX$ systems [2-11]. Properties characteristic of high-transition temperature (high-$T_C$) superconductors have been observed, notably the relatively low density of superconducting carriers produced by doping [12,13], evidence for quasi-two-dimensional superconductivity from diamagnetism [14] and critical fields [8,10,14,15], weak isotope effect in $T_C$ [16, 17], and high $T_C$ when compared to the cubic metal-$M$ nitride [18]. Reviews of these materials are presented in [12] and 13.

Distinguished as to $[MNX]_2$ structure and layering symmetry, $A_x(S)_yMNX$ compounds form in two principle crystal structures denoted as α and β. More extensive research is available for superconducting β-$A_x(S)_yM$NCl compounds based on Zr and Hf for $M$, where especially useful data have been obtained for compositions over a range of concentrations x of alkali, alkaline earth and rare earth or lanthanide elements for $A_x$ and also with various co-intercalants $(S)_y$ [7-9,13,19,20]. From such studies, this work identifies for a given compound an optimal value $x_{opt}$ of cation concentration and an optimal transition temperature $T_{C0}$. This finding mirrors a basic attribute of high-$T_C$ superconductors, where optimal doping and stoichiometry yields highest transition temperature $T_{C0}$. Experiments comparing various β-$A_x(S)_yM$NCl compounds indicate $T_C$ tends to be correlated with the transverse distance $d$ between the $[MNCl]_2$ blocks (basal-plane spacing) [7-9].

For α-$A_x(S)_yTiNX$ compounds, the correspondence between $T_C$ and $d^{-1}$ is nearly linear, which has been interpreted as strong evidence of variable interlayer coupling [10,11], and quantitatively treated accordingly [21,22]. Analysis of the data for $X$ = Cl compounds in [10] showed that pair-breaking from remote Coulomb scattering (RCS) by $A_x$ cations leads to depression of the observed $T_C$ relative to the optimal $T_{C0}$ calculated from the interlayer Coulomb interaction model of high-$T_C$ superconductivity [21,23]. In arguing against phonon-mediated superconductivity, other theoretical ideas considered in [10] include the RVB high-$T_C$ theory of the cuprates [24,25], a multilayer conductor model applied to $Li_{0.48}(THF)_yHfNCl$ (THF is tetrahydrofuran, $C_4H_8O$) [26,27], band structure [28,29], spin and charge fluctuations in TiNCl [29], and a pair-hopping mechanism applied to $K_{0.25}TiNCl$ [30]. Indeed, many different approaches specific to high $T_C$ have been advocated [31]. Superconductivity in doped $MNX$ systems presents an understandably challenging problem [32], one not very conducive to most of these theories that nominally assume an optimal superconductive state.

As applied to the $A_x(S)_yMNX$ compounds, the interlayer interaction model requires identifying the two types of layered charge reservoir structures [21,23]. In this case, type I is assigned to the $[MNX]_2$ structure, which hosts and sustains the superconducting current, and the intercalant layer $A_x(S)_y$ is designated as type II (including $A_x$-only intercalation at y = 0), providing the mediation for the superconductive pairing interaction. Coupling occurs between adjacent ionic layers, thus involving the outer halogen layers of $[MNX]_2$ and the $A_x$ cations in the intercalant layers, separated by a transverse distance ζ. The optimal transition temperature $T_{C0}$ occurs upon the formation of participating charges in the two reservoirs for x at optimal doping $x_{opt}$.

The existence of two interacting reservoirs in these materials is clearly evidenced by considering the difference between $d$ and $\zeta$, where the former is always defined and non-zero, and the latter is unrealized in the absence of two physically separated and adjacent interacting layers. This key distinction explains why the correlation of the measured $T_C$ with $d^{-1}$ breaks down for $Li_{0.13}TiNCl$ [10], indicating that the length $d$ is not the relevant length parameter involved [21,22]. Since the Li cations occupy sites between the Cl anions [10,22,28], a spatially separated intercalation layer is not formed; hence, $Li_{0.13}TiNCl$ does not possess the requisite two-layer interaction structure for superconductivity from interlayer Coulombic interactions. The fact that $Li_{0.13}TiNCl$ (for which $\zeta$ is unrealized) does not behave in the same manner as the other seven $\alpha$-TiNCl-based compounds of [10] provides strong support for pairing based on interlayer Coulomb interactions (the superconductivity observed for $Li_{0.13}TiNCl$ is most likely related to phonon-mediated TiN with $T_C = 5.6$ K [10,18]). The absence of superconductivity in $H_xZrNCl$ [33], where in this case the H impurities likely occupy the 6c site between the Zr-N and Cl ions and dope the type I reservoir, and in Li-doped $\alpha$-HfNBr is also thusly explained [22]. Zhang *et al.* [10] also compare their results with earlier studies of intercalated Bi-based cuprates [34], in which intercalation of charge-neutral molecules between the double BiO layers leaves $T_C$ unchanged. Since $\zeta$ for the Bi-based cuprates is defined as the distance between adjacent SrO and $CuO_2$ layers [21], which doesn't change upon such intercalation, this behavior is expected and confirms that $\zeta$, not $d$, is the length which governs $T_{C0}$ in high-$T_C$ superconductors. From this, one can conclude that the superconductivity exhibited by the remaining $A_x(S)_yTiNCl/Br$ compounds considered in [10] and [11] clearly originates from interlayer Coulomb interactions, which should also be true for all members of this superconductor family.

These compounds are unique among high-$T_C$ materials, since the mediating layer adjacent to the superconducting condensate is incomplete, containing a high density of vacancies. Even absent any disorder in the $[MNX]_2$ structures [35], close proximity of disordered $A_x$ cations can cause RCS in a manner analogous to the observed suppression of carrier mobility by remote fixed charges in two-dimensional systems [36]. For the $\alpha$-form Ti-based compounds [10,11], in which large resistivities at $T_C$ and broad resistive transitions $\Delta T_C$ provide evidence of significant scattering, it is found that RCS pair breaking has the effect of limiting the superconductivity from achieving an optimal state for $d < 20$ Å, causing $T_C < T_{C0}$. Given the interlayer Coulomb interaction model and compound structure, the $T_C \propto d^{-1}$ dependence observed for $A_x(S)_yTiNCl/Br$ [10,11] is readily explained, including the anomalous behavior of $Li_{0.13}TiNCl$ which does not follow the linear trend. RCS pair breaking appears significantly reduced in the $\beta$-form materials with Zr or Hf for $M$ that exhibit comparatively sharper superconducting transitions.

Section 2 presents a general description of the interlayer interaction model for $T_{C0}$ at optimal doing and a model for pair breaking from RCS. Section 3 explains the model's application to interpreting experimental $T_C$ in $A_x(S)_yMNX$ systems, and concluding remarks are given in Section 4.

## 2 Model $T_C$

Superconductivity in $A_x(S)_yMNX$ compounds is treated according to the authors' general interlayer coupling model for calculating the optimal transition temperature $T_{C0}$ in high-$T_C$ superconductors, which was previously shown to yield accurate results for $T_{C0}$ when compared with experiment for 39 optimal compounds, e.g., cuprates [23,37], iron pnictides [23], iron chalcogenides [38], and other superconductor families [23,39]. In these prior studies of high-$T_C$ superconductors, optimal $T_{C0}$ has been identified experimentally as the composition exhibiting maximal $T_C$ and Meissner-effect fraction, and minimal superconducting transition width $\Delta T_C$. Significant observations in α-$A_x(S)_y$TiN$X$ compounds, distinguishing them from the β-form compounds and typical high-$T_C$ superconductors, are marginally metallic electrical conductivity, broad $\Delta T_C$, and rounded superconducting onsets. To account for these features of the superconducting transition, an RCS model has been introduced to quantify the effect of pair-breaking interactions that force $T_C$ below $T_{C0}$ [21].

## 2.1 Calculation of Optimal $T_{C0}$

The $A_x(S)_yMNX$ compounds are modeled as alternating layered structures, comprising the $[MNX]_2$ structural block containing the superconducting charges and the neighboring intercalant layer $A_x(S)_y$ containing the mediating charges [21]. These are denoted in the model as the type I and type II charge reservoir layers, respectively. Optimal doping suggests equilibrium between the two reservoir types. The superconducting transition temperature is spatially dependent upon the indirect Coulomb interaction across the transverse distance $\zeta$, which is measured between the outer layer halogen $X$ ions in the $[MNX]_2$ block and the locus of the doping charge in the intercalation layer. The layered structure of $A_x(S)_yMNX$ is characterized by a transverse spacing $d$ between $[MNX]_2$ blocks of thickness $d_2$, such that the intercalant thickness is determined by $d - d_2$ [13]. Taking the locus of doping charge as the intercalant-layer midplane, the distance $\zeta$ is determined as one-half the intercalant thickness according to,

$$\zeta = (d - d_2)/2 \ . \tag{1}$$

Equation (1) is used to determine $\zeta$ for compounds where the cations $A_x$ occupy the midplane and co-intercalant molecules, when present, are uncharged and unpolarized. In cases of covalent bonding of the intercalated cations with the co-intercalated molecules $(S)_y$, $\zeta$ is determined by the $c$-axis projection of the $(A-S)^+/X^-$ dipole length instead of Eq. (1). As previously noted [21], the $d_2$ for the α-$A_x(S)_y$TiNCl compositions of [10] are approximately the same as for pristine α-TiNCl [13], yielding significant correlation between the functional dependences of $T_C$ on $d$ and $\zeta$. On the other hand, $d_2$ for the compounds derived from the β-forms ZrNCl and HfNCl can vary significantly. Structural and superconductivity data for the $A_x(S)_yMNX$ compounds considered herein are presented in Tables 1 and 2. In each case, the interlayer interaction length is the shorter distance $\zeta$ determined from Eq. (1), rather than the spacing $d$, using directly measured values of $d_2$ when available.

Calculated optimal transition temperatures for $A_x(S)_yMNX$ compounds are obtained from the general expression previously applied to high-$T_C$ superconductors of optimal compositions,

$$T_{C0} = k_B^{-1} \ \beta \ \zeta^{-1} \ (\sigma\eta/A)^{1/2} \ , \tag{2}$$

where the universal constant $\beta = 0.1075(3)$ eV-Å$^2$ was determined by fitting Eq. (2) to experimental data for $T_{C0}$ [23]. The quantity $\sigma\eta/A$ is the two-dimensional density of interaction charges given by the model. Determined per formula unit by doping as discussed below, σ is the participating fractional charge in the type I reservoir, $A$ is the crystal basal plane area, and η is the number of charge

**Table 1.** Structural and electronic parameters of optimal β-form $A_x(S)_yM$NCl compounds for measured transition temperature $T_C$, interlayer distance $d$, $[M$NCl$]_2$ thickness $d_2$, basal plane area $A$, superconducting onset value $x_0$, interaction distance $\zeta$, and modeled optimal $T_{C0}$.

| Compound | $T_C$ (K) | $d$ (Å) | $d_2$ (Å) | $A$ (Å²) | $x_0$ | $\zeta$ (Å) | $T_{C0}$ (K) |
|---|---|---|---|---|---|---|---|
| Li$_{0.08}$ZrNCl [19] | 15.1 | 9.3367 | 6.21 [a] | 11.286 | 0.05 [13] | 1.5634 | 14.54 |
| Li$_{0.13}$(DMF)$_y$ZrNCl [7] | 13.7 | 13.01 | 6.21 [a] | 11.3233 [b] | 0 | 3.400 | 13.90 |
| Na$_{0.25}$HfNCl [3,13] | 24 | 9.8928 | 6.58 | 11.1484 | 0.15 [13] | 1.656 | 25.22 |
| Li$_{0.2}$HfNCl [20] | 20 | 9.40 | 6.21 [a] | 11.1195 | 0.14 | 1.595 | 20.31 |
| Li$_{0.2}$(NH$_3$)$_y$HfNCl [20] | 22.5 | 12.10 | 6.58 [a] | 11.1117 [c] | 0 | 2.76 | 21.44 |
| Ca$_{0.11}$(NH$_3$)$_y$HfNCl [9] | 23 | 12.05 | 6.58 [a] | 11.1251 | 0 | 2.735 | 22.68 |
| Eu$_{0.08}$(NH$_3$)$_y$HfNCl [8] | 23.6 | 11.914 | 6.58 [a] | 11.1117 | 0 | 2.667 | 24.30 |
| Li$_{0.2}$(THF)$_y$HfNCl [20] | 25.6 | – | – | 11.1117 [c] | 0 | (2.37) | 24.97 |
| Ca$_{0.11}$(THF)$_y$HfNCl [9] | 26 | 15.0 | – | 11.1117 [c] | 0 | (2.37) | 26.18 |

Values for $\zeta$ in parentheses are speculative. Numeric footnotes correspond to cited references.
[a] Host or related material value is used in absence of refinement data.
[b] Assumed from Li$_{0.13}$ZrNCl in [19], Fig. 1(c).
[c] Assumed same as Eu$_{0.08}$(NH$_3$)$_y$HfNCl

**Table 2.** Structural and electronic parameters of α-form $A_x(S)_y$TiN$X$ compounds for measured transition temperature $T_C$, interlayer distance $d$, $[$TiN$X]_2$ thickness $d_2$, basal plane area $A$, interaction distance $\zeta$, modeled optimal $T_{C0}$, pair-breaking parameter $\alpha$, and hypothetical optimal transition temperature $T_{C0}^{(\alpha\to0)}$, calculated with pair-breaking effect removed using model of $T_C$ for $A_x(S)_y$TiNCl.

| Compound | $T_C$ (K) | $d$ (Å) | $d_2$ (Å) | $A$ (Å²) | $\zeta$ (Å) | $T_{C0}$ (K) | $\alpha$ (meV) | $T_{C0}^{(\alpha\to0)}$ (K) |
|---|---|---|---|---|---|---|---|---|
| Na$_{0.16}$(PC)$_y$TiNCl [10] | 7.4 / 6.3 | 20.53 | 5.183 [a] | 13.0331 | 7.6735 | 6.37 | 0 | 7.53 |
| Na$_{0.16}$(BC)$_y$TiNCl [10] | 6.9 | 20.7435 | 5.183 [a] | 13.0331 [b] | 7.7803 | 6.28 | 0 | 7.03 |
| Na$_{0.16}$(THF)$_y$TiNCl [10] | 10.2 | 13.105 | 5.183 [a] | 12.9753 | 3.9610 | 12.36 | 0.230 | 12.21 |
| Li$_{0.13}$(THF)$_y$TiNCl [10] | 9.5 | 13.0012 | 5.183 [a] | 13.1277 [d] | 3.9091 | 11.23 | 0.184 | 11.19 |
| Na$_{0.16}$TiNCl [10] | 18.0 | 8.442 | 5.150 [c] | 13.1564 | 1.6460 | 29.55 | 1.175 | 29.35 |
| K$_{0.17}$TiNCl [10] | 17.0 | 8.77884 | 5.182 [c] | 13.3720 [c] | 1.7984 | 27.65 | 1.085 | 27.80 |
| Rb$_{0.24}$TiNCl [10] | 16.0 | 9.21038 | 5.000 [c] | 13.2830 [c] | 2.1052 | 28.16 | 1.224 | 28.34 |
| Na$_{0.21}$(PC)$_y$TiNBr [11] | 8.6 | 15.7 | 5.499 [a] | 13.5137 [a] | 5.1005 | 10.78 | 0.231 | 9.74 |
| Na$_{0.21}$(THF)$_y$TiNBr [11] | 11.0 | 12.9 | 5.499 [a] | 13.5137 [a] | 3.7005 | 14.86 | 0.404 | 14.23 |
| Li$_{0.37}$(THF)$_y$TiNBr [11] | 7.5 | 14.186 | 5.499 [a] | 13.4312 | 4.3435 | 16.85 | 0.910 | 11.14 |
| Na$_{0.23}$TiNBr [11] | 15.2 | 8.942 | 5.499 [a] | 13.5137 | 1.7215 | 33.42 | 1.778 | 31.17 |
| K$_{0.21}$TiNBr [11] | 17.2 | 9.464 | 5.499 [a] | 13.5951 | 1.9825 | 27.65 | 1.067 | 28.92 |
| Rb$_{0.22}$TiNBr [11] | 16.3 | 9.691 | 5.499 [a] | 13.5844 | 2.0960 | 26.78 | 1.066 | 27.62 |

[a] Host or related material value is used in absence of refinement data. Numeric footnotes correspond to cited references.
[b] Assumed from Na$_{0.16}$(PC)$_y$TiNCl.
[c] $d_2$ assumed from Na$_{0.22}$TiNCl, K$_{0.22}$TiNCl, and Rb$_{0.19}$TiNCl in Table 7 of [13]; $A$ assumed from Table 1 of [28].
[d] Assumed from Li$_{0.13}$TiNCl.

carrying layers in the type II reservoir type II reservoir and given by $\eta = 1$ for $A_x(S)_y$.

Writing $\beta = e^2\Lambda$ in Eq. (2), one finds the interlayer Coulomb potential in the energy scale as $e^2\zeta^{-1}$, with a length scale $\Lambda = 0.00747$ Å equal to about twice the reduced electron Compton wavelength. Equation (2), which is confirmed by experiment, follows from an interaction term in the Hamiltonian containing the interlayer Coulomb potential $V_{int}(q) \sim \exp(-q\zeta)$, where $q$ is the wave vector. The mediating bosons are assumed to be electronic excitations, such as those considered within a model of multiple charge layers [26,27]. The coupling strength is calculated from the quadrature average of the interaction force as $\lambda \propto \langle|qV_{int}(q)|^2\rangle \propto e^4(\ell\zeta)^{-2}$, where the result derives from a real-space approximation using an interaction charge density defined as $\ell^{-2} = \sigma\eta/A$ [23]. Applying strong coupling in the form $T_C \propto \lambda^{1/2}$ [40-42], one arrives at the right side of Eq. (2) apart from a determination of $\beta$. The $T_{C0}$ defined by Eq. (2) should be considered as an upper limit on the experimentally observed transition temperature, given $T_C < T_{C0}$ for non-optimal materials (see, e.g., [37]).

In expressing the central tenet of the model, the participating charge fraction $\sigma$ entering Eq. (2) is determined for optimal materials from the difference between the optimal dopant charge stoichiometry $x_{opt}$ and the minimum stoichiometric value $x_0$ required for superconductivity. Well studied examples are $x_{opt} = 0.163$ taken relative to $x_0 = 0$ in $La_{1.837}Sr_{0.163}CuO_{4-\delta}$ and $x_{opt} = 6.92$ relative to $x_0 = 6.35$ in $YBa_2Cu_3O_{6.92}$. This definition requires an insulating end material and, since all three undoped host $MNX$ materials ($M$ = Ti, Zr, and Hf) discussed herein are essentially insulating [13,43-45], it is also applicable to the associated $A_x(S)_yMNX$ superconductors. Compounds with $M$ = Ti and co-intercalated compounds with y >

0 have $x_0 = 0$; compounds with $M$ = Zr or Hf and y = 0 have $x_0 \geq 0$ [13,46].

Direct doping may be either cationic or anionic, occurring in the type I reservoir as in the case of $La_{2-x}Sr_xCuO_{4-\delta}$, the type II reservoir, e.g., $Ba_2Y(Ru_{1-x}Cu_x)O_6$ [23], or in both as in the ternary Fe-based chalcogenides (e.g., $A_xFe_{2-y}Se_2$) [38] and $(Ca_xLa_{1-x})(Ba_{1.75-x}La_{0.25+x})Cu_3O_y$ [37]. For $A_x(S)_yMNX$ (excluding $X$-deficient materials), doping is introduced by $A_x$ intercalation such that $\sigma$ is determined by doping in only the type II reservoir according to the simplified relation,

$$\sigma = \gamma \left[v \left(x_{opt} - x_0\right)\right], \tag{3}$$

where $x_{opt}$ is the optimal x for dopant species $A_x(S)_y$, and $x_0$ is the threshold value of x for superconductivity. The optimal $x_{opt}$ is determined experimentally as shown in Section 3. Equation (3) is an application of the general form for doping in both reservoir types [38,39]. The factor $v$ is the charge state of the $A_x$ cations taken to equal their valence. The $\gamma$-factor is derived from considerations of charge allocation within a given structure. Generally applied to all optimal high-$T_C$ superconductors, the procedure assumes the dopant charge to be shared equally between the charge reservoirs, and further distributed pair-wise between like charge-carrying layers within the reservoirs. Consequently, $\gamma$ can be determined by applying the following two charge allocation rules [23]:

(1a) Sharing between N (typically 2) ions or structural layers introduces a factor of 1/N in $\gamma$.

(1b) Doping is shared equally between the two reservoirs, resulting in a factor of 1/2.

For the $A_x(S)_yMNX$ compounds, determination of $\gamma$ is somewhat comparable to that of $(Ba_{0.6}K_{0.4})Fe_2As_2$ [23], for which a structural analogy was previously noted [12]; co-intercalated molecules are assumed to contribute no doping charge. Application of rule (1b) contributes a factor 1/2 to $\gamma$, where the charge is divided equally between the two reservoirs. Two applications of rule (1a) contribute a factor (1/2)(1/2) to $\gamma$, first applied pair-wise between the $X$ and $M$-N layers, and then between $M$ and N. Thus we have $\gamma = (1/2)(1/2)(1/2) = 1/8$ in Eq. (3). Since $\gamma$ less than unity is thusly obtained, the participating charge fraction $\sigma$ is generally smaller than the doping content associated with $x_{opt}$. For optimal cuprate compounds where the doping is not known, $\sigma$ may be calculated by scaling to $YBa_2Cu_3O_{6.92}$, as discussed in [23,37], and [39].

## 2.2 Effect of Disorder on $T_C$

Significantly broadened superconducting transitions have been reported for $\alpha$-form compounds with $M$ = Ti, which are taken as evidence of disorder influencing observed $T_C$. Magnitudes of broadening are determined from resistance-$vs.$-temperature curves as $\Delta T_C = (T_{peak} - T_{mid})$, where $T_{peak}$ corresponds to the maximum resistance just above or at $T_C$ and $T_{mid}$ is the transition midpoint. Available data give $\Delta T_C/T_{peak}$ of 0.29, 0.48, and 0.64 for $Na_{0.16}(S)_y TiNCl$ compounds with $y = 0$ and $y > 0$ for $S$ = THF or PC (polypropylene carbonate, $C_4H_6O$), respectively [10], and $\Delta T_C/T_{peak} = 0.30$ for $K_{0.21}TiNBr$ [11]. Large broadening is also reflected in rounded transition onsets observed in magnetic susceptibility [10,11]. Although resistivities exceeding the Ioffe-Regel limit are indicated, the data are suspected to include extrinsic effects arising from grain boundaries [10,11].

In contrast, $\beta$-forms with $M$ = Zr and Hf display significantly sharper resistance transitions; e.g., $\Delta T_C/T_{peak} \approx 0.06 - 0.09$ for $Li_{0.08}ZrNCl$ [19,47] and 0.04 for $Li_{0.48}(THF)_y HfNCl$ [35]. These indicators of more ordered superconducting transitions in $\beta$-forms are consistent with the minimal disorder scattering in $\beta$-$Li_x ZrNCl$ deduced from magneto-transport measurements, where an analogy was drawn to modulation-doped semiconductor quantum wells [47].

The presence of disorder in $\alpha$-$A_x(S)_y TiNCl$ compounds was previously treated as an intrinsic property and attributed to RCS [21]. Since the $\alpha$-form $[TiN]_2$ structures have smaller thickness $d_2$, when compared to their $\beta$-form counterparts, the dopant ions $A_x$ act more effectively as Coulomb scattering centers, because they lie in closer proximity to the $[TiN]_2$ superconducting channel. In principle, the strength of Coulomb scattering is diminished by screening [36], which is more efficient in $\beta$-form compounds comprising the species Zr and Hf, owing to their higher atomic numbers.

The effect of disorder was previously treated as inducing pair-breaking and depressing the observed $T_C$ relative to the optimal $T_{C0}$ of Eq. (2) [37,48,49]. As applied to the charge-compensated cuprate $(Ca_xLa_{1-x})(Ba_{1.75-x}La_{0.25+x})$-$Cu_3O_y$, pair-breaking scattering is associated with resistivity and accounts for the implicit dependence of $T_C$ on doping x [37]. As applied to $\alpha$-$A_x(S)_y TiNX$, the pair-breaking is associated with RCS [21]. For these materials, the pair-breaking formalism is modeled by the expression,

$$\ln (T_{C0}/T_C) = \psi( \tfrac{1}{2} + \alpha/2\pi k_B T_C ) - \psi( \tfrac{1}{2} ) \quad (4)$$

where $\psi$ is the digamma function and $\alpha$ is the pair-breaking parameter.

For treating pair-breaking in $\alpha$-$A_x(S)_y TiNX$, $\alpha$ is modeled by considering the spin-orbit scattering associated with the ionized $A_x$ dopants. Although pair-breaking by spin-orbit scattering from random impurities is usually regarded as a minor perturbation [50], particularly for nodeless pairing symmetries as is suggested for $A_x(S)_y MNX$ superconductivity [14,15,45] (and substantially enhanced for d-wave [51] or noded symmetries [50]), it does appear to be significant in $\alpha$-$A_x(S)_y$TiN$X$ compounds owing in part to their layered structure. Adopting the analogy with modulation doped quantum wells [47], the carriers in the superconducting [TiN$X$]$_2$ channel are scattered by fixed charges from $A_x$ cations in the intercalation layer. A prior study [21] of $\alpha$-$A_x(S)_y$TiNCl compounds from [10] showed that $\alpha$ mimics the form factor for the indirect Coulomb potential $\exp(-qz)$, which is expressed by wave vector q and transverse distance z between channel and scattering sites [36], finding z is given by $\zeta$ and $q^{-1}$ approximately by an *ab* plane lattice spacing [21].

In modulation-doped semiconductor structures, spin-orbit scattering rates have been determined in closed form by assuming a white noise distribution for the scattering potential from random impurities in the doping layers [52]. In the case of the $\alpha$-$A_x(S)_y$TiN$X$ compounds, the $A_x$ cations occupy a fraction ~2x of the available sites (assuming full occupancy at x = 1/2) constrained to an intercalation layer sheet, although site occupancy distributions are evidently undetermined from X-ray refinement analysis [10,11,28]. The $A_x$ scattering centers are therefore more ordered than impurities in a doping structure of extended thickness, and one may assume that fluctuations in the RCS field are dominated by wave vectors on the order of $A^{-1/2}$. Hence, for the unit valency alkali intercalants $A_x$ of $A_x(S)_y$TiN$X$ in [10] and [11], $\alpha$ is modeled in terms of two empirical parameters as,

$$\alpha = a_1 x \exp(-k_1 \zeta) , \qquad (5)$$

where the parameter $k_1$ models the dominantly large-$q$ scattering, $\zeta$ is the RCS interaction distance, and $a_1$ represents the strength of the random RCS field, which scales with dopant content x. This form is consistent with spin-orbit scattering being small compared to scalar impurity scattering, since the rate $2\hbar^{-1} a_1$ associated with the prefactor in Eq. (5) was found to be small compared to the transport scattering rate $\tau^{-1}$ estimated from resistivity [21].

For the $A_x(S)_y$TiNCl compounds in [10], the transition temperatures $T_C$ determined from Eqs. (2), (4) and (5) have the tendency to scale as $\zeta^{-1}$ [21], which, as noted above from Eq. (1), yields a correlation between $T_C$ and $d^{-1}$ for the $\alpha$-form compounds, similar to the linear trends reported by Zhang et al. [10,11,22]. In Section 3.3, results for $\alpha$ derived from data on $A_x(S)_y$TiN$X$ with $X$ = Cl and Br are compared with Eq. (5).

## 3 Experimental $T_C$

In order to understand the experimental results for $T_C$ in the $A_x(S)_y$$M$N$X$ system, one must first consider the transition temperature $T_{C0}$ for optimal compounds defined by combining Eqs. (2) and (3) with $\gamma$ = 1/8 and x at optimal cation content $x_{opt}$,

$$T_{C0} = k_B^{-1} \beta \zeta^{-1} [(1/8) \, v(x_{opt} - x_0) \, /A]^{1/2} . \qquad (6)$$

The model picture adopted assumes that the superconducting condensate is hosted by the [$M$N$X$]$_2$ structure and mediated by the charges in the $A_x(S)_y$ intercalant layer. Although the pairing interaction occurs between electrons in the halide ($X$) layers and the presumably more localized charges associated with cation intercalation layers, the detailed structure of the $M$N$X$ host and screening from the halide and intercalated cation layers determine $T_{C0}$ and $T_C$.

Data for 11 intercalated $\beta$-form compounds are studied, four based on $\beta$-ZrNCl/Br and seven based on $\beta$-HfNCl. As indicated in Section 2.2, these compounds appear to exhibit little or no RCS pair breaking effects, unlike the $\alpha$-TiN$X$-based materials, which is attributed in part to differences in basal-plane symmetries [13] and possibly also to greater occupancy ordering of the intercalated charge species. Not surprisingly, these materials also possess the highest $T_C$s among the $A_x(S)_yMNX$ superconductors; seven β-form compounds with $T_C \approx T_{C0}$ and $\zeta$ defined as in Eq. (1) are noted. Section 3.1 presents analysis of the $A_x(S)_y$ZrNCl/Br system, with particular attention paid to determining the optimal doping $x_{opt}$ and establishing non-zero $x_0$ from experiment. Section 3.2 discusses the $A_x(S)_y$HfNCl compounds, in particular observations of $T_C$ weakly varying with dopant concentration and indications of delocalized $A$-$S$ charge. In Section 3.3, it is shown that by using this model for $T_{C0}$ in conjunction with the pair-breaking expression of Eq. (4), one may understand the variation of $T_C$ with $d$ for α-$A_x(S)_y$TiNCl/Br as reported in [10] and [11]. Since $\zeta$ depends functionally on $d$, correlations between $T_C$ and $d$ are thus possible within $A_x(S)_y$TiN$X$ systems having presumably small variations in $d_2$.

From consideration of the $T_C$ values, $d$-spacings and anisotropy ratios of the upper critical field $H_{C2}$, coupled with the $c$-axis coherence distances $\xi_c$ of Na$_{0.16}$TiNCl and Na$_{0.16}$(THF)$_y$TiNCl (given in [10]), the superconducting volume in $A_x(S)_y$TiNCl is indicated to comprise the complete [TiNCl]$_2$ block and intercalant layer $A_x(S)_y$ [21]; this finding is assumed generally true for the component structures of all $A_x(S)_y$[$MNX$]$_2$ materials. The interaction distance $\zeta$ is written in Eq. (1) as a function of $d$ and $d_2$ for consistency and to reflect established notation. Whenever possible, actual structural refinement data is used for $d_2$; absent such information, the value of a related material or that of the host $MNX$ (see Fig. 4 and Table 4 of [13]) is substituted.

### 3.1 β-$A_x(S)_y$ZrNCl

The β-$A_x(S)_y$ZrNCl compounds merit special attention for testing Eq. (6), owing to the availability of numerous experimental measurements, particularly the dependencies of material properties on cation dopant concentration x. The well-studied and consequently exemplary compound is Li$_x$ZrNCl, for which optimal stoichiometry is identified as occurring at $x_{opt}$ = 0.08 at which point the full superconducting volume is reached and $T_C$ (= 15.1 K) is maximized [19,20]. Underdoping occurs for x < 0.08, as identified by diminished superconducting volumes that vanish for x < 0.05 [19], indicating $x_0$ = 0.05 is the minimum value of x for superconductivity. Muon-spin-depolarization rates $\sigma_\mu$(T→0) for Li$_x$ZrNCl are linear in (x − $x_0$) with $x_0$ = 0.05 obtained by extrapolation [51]; also, plasma frequency data show that $\omega_p^2$ extrapolates to zero for x ≈ $x_0$ [46]. Signature characteristics of overdoping occur for x > $x_{opt}$ in that $T_C$ is a decreasing function of x, falling off precipitously from 15.1 K for x between 0.08 and 0.2 and reaching a minimum of about 11.5 K for 0.2 < x < 0.4 [19,47]. Other studies of Li$_x$ZrNCl for x equal to 0.13 [7], 0.16 [1,5,13], and 0.2 [6,13] have also reported lower values of transition temperatures in the overdoped region x > $x_{opt}$. Overdoping is observed in the superconducting state through $\sigma_\mu$(T→0) and in the normal state through $\omega_p^2$, both of which monotonically increase with x for x > 0.1, notably opposite to the trend in $T_C$ vs. x [46,51]. This negates the notion of the superfluid density being a dominant factor in determining $T_C$ in the overdoped regime, as suggested elsewhere [15]. A related compound Li$_{0.15}$ZrNBr with $T_C$ = 12 K has also been studied [4], which is evidently an analogue of overdoped Li$_{0.16}$ZrNCl with $T_C$ = 12.5 K [1,13].

Continuous incorporation of Li and charge doping have been concluded from an x dependence in certain Raman modes, particularly for in-plane vibrations of the [ZrNCl]$_2$ block (mode denoted "A" in [19]).

Within this x variation, the underdoping and overdoping regimes appear to be reflected in the x-dependence of the lattice parameters [19]. The lattice parameter $c$ is particularly sensitive to the intercalant thickness and dictated to a significant extent by the Li-Cl bond length that determines $\zeta$ according to Eq. (1), since $d$ is given by $c/3$ while the $[ZrNCl]_2$ thickness $d_2$ tends to show little variation with intercalation. The variation of $c$ *vs.* x is strongest at low x and crosses over to a weaker dependence for x greater than about 0.1, which is rather close to the optimal doping point $x_{opt}$. This behavior suggests that participating charges are introduced into the interaction layers for $x \leq x_{opt}$ with the equilibrium charge structure of the Li-Cl interaction layers fully formed at $x = x_{opt}$. For $x > x_{opt}$ the excess charges are transferred to the $[ZrNCl]_2$ reservoir, yielding weak variation of intercalant thickness with the further increases in x. This non-participating charge fraction, being more localized in the $[ZrN]_2$ structures and minimally affecting the Cl sites, is assumed benign with respect to the superconducting pairing interaction. A similar change in slope is revealed in the careful measurements of $\omega_p^2$ *vs.* x, which nearly coincide with results from first principles calculations of band structure [46]. In addition, the damping $\hbar\tau^{-1}$ is reported to be greater for $x > x_{opt}$ containing non-participating charge.

Data for the optimal compound $Li_{0.08}ZrNCl$ [19] are presented in Table 1, where $d_2$ of the unintercalated host ZrNCl is used for calculating $\zeta = 1.5634$ Å from Eq. (1). With $\gamma = 1/8$, $v = 1$, and $x_0 = 0.05$ [13], Eq. (3) yields $\sigma = 0.00375$ and from Eq. (6) the calculated $T_{C0} = 14.54$ K. Figure 1 shows the measured $T_C$ plotted against the factor $\zeta^{-1}[(1/8) v(x-x_0) /A]^{1/2} = 1.1660$ nm$^{-2}$ along with a diagonal line representing the model function $(x = x_{opt})$ for $T_{C0}$ in Eq. (6). The close proximity of calculated $T_{C0}$ to measured $T_C$ substantiates the applicability of the model to the $x_{opt} = 0.08$ composition. Also reported in [35] is the compound $Zn_{0.04}ZrNCl$ with $T_C = 15$ K, which is possibly a divalent doping analogue of $Li_{0.08}ZrNCl$.

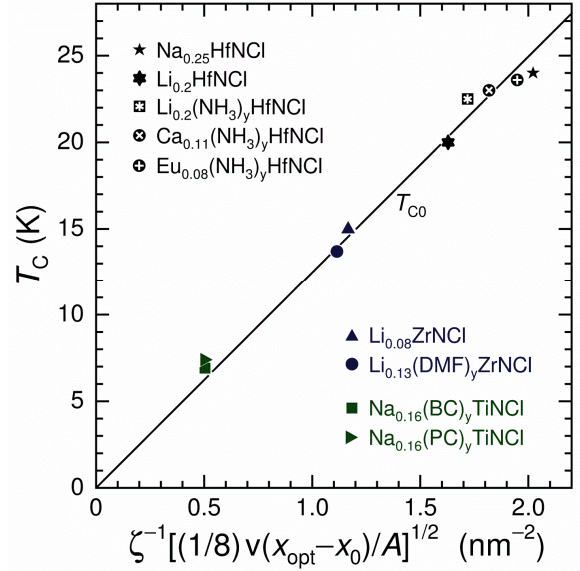

**Fig. 1** Plot of measured transition temperature $T_C$ vs. $\zeta^{-1}[(1/8)v(x_{opt}-x_0)/A]^{1/2}$ for the high-$T_C$ compounds listed in Table 1 and compounds with $\alpha = 0$ from Table 2. Interaction distance $\zeta$ is from Eq. (1), $v(x_{opt} - x_0)$ is the optimal doping $x_{opt}$ relative to the superconductivity onset value $x_0$ and multiplied by valence v, and $A$ is basal area per formula unit. The *line* represents $T_{C0}$ for optimal compounds from Eq. (6).

Table 1 and Fig. 1 include data for the co-intercalated compound $Li_{0.13}(DMF)_yZrNCl$ (DMF is $(N,N)$-dimethylformamide, $C_3H_7NO$) with measured $T_C = 13.7$ K [7]. The indicated $T_{C0} = 13.90$ K is calculated from $\zeta^{-1}[(1/8) v(x-x_0) /A]^{1/2} = 1.1142$ nm$^{-2}$ with the assumption $x_0 = 0$ inferred from [13]; the close agreement of model with experiment indicates nearly optimal stoichiometry. A somewhat related compound $Eu_{0.16}(NH_3)_yZrNCl$ with $T_C = 12.7$ K, $d = 11.889$ Å, and $A = 11.2943$ Å$^2$ has also been studied [8]. With $\zeta = 2.8395$ Å from Eq. (1) and $v = 3$ for $Eu^{+3}$, one obtains a substantially greater $2.5669$ nm$^{-2}$ value for $\zeta^{-1}[(1/8) v(x - x_0) /A]^{1/2}$, from which the model indicates an overdoped composition.

## 3.2 β-A$_x$(S)$_y$HfNCl

Like β-Li$_x$ZrNCl, compositions of the β-Li$_x$HfNCl compound also form solid solutions for $0.1 \leq x \leq 0.5$, as deduced from observations of x-dependent Raman shifts [20]. By interpolating the x dependence in superconducting volume fractions and transition temperatures [20], the minimum Li content for superconductivity, $x_0 = 0.14$, is estimated. The onset point for maximum superconducting volume fraction (~97%) together with the highest $T_C$ (20.0 K) occur for $x_{opt} = 0.2$; values of $T_C$ approaching 20 K and 97% volume fraction are found in the higher doping range $0.2 < x \leq 0.5$ [20]. The $x_{opt}$ point is independently shown in the variation of $c$ (and $d = c/3$) vs. x, which exhibits a crossover from a strong to a weak x-dependence at $x \approx 0.2 = x_{opt}$ [20], analogous to findings for Li$_x$ZrNCl [19], and indicating the presence of interaction charges in the Li and Cl layers for $x \leq x_{opt}$. In the region $x > x_{opt}$, transfer of non-participating excess charges to the [HfNCl]$_2$ reservoir is indicated by the slight variations in $T_C$ and volume fraction, weak variations in lattice parameters, and continuous shifts in frequencies of vibrations parallel to the conducting planes and the plasma frequency $\omega_p$, for $0.3 \leq x \leq 0.5$ [20], which are consistent with band structure calculations of $M$-related orbital filling upon doping [53,54]. With $\sigma = (0.2 - 0.14)/8$ determined for the participating charge fraction, the calculated optimal $T_{C0} = 20.31$ K is obtained for the optimal Li$_{0.2}$HfNCl compound; the parameters are listed in Table 1 and provide the associated datum point in Fig. 1.

Superconductivity in Na$_x$HfNCl, which is observed over a comparatively limited range in x, exhibits highest reported $T_C = 24$ K for $x = 0.25$ [3]; $T_C$ decreases for higher doping ($T_C = 22$ K at $x = 0.28$; $T_C = 19$ K at $x = 0.29$) [2], attributable primarily to a combination of increasing ζ and overdoping for $x > x_{opt} \approx 0.25$. The relevant parameters for the optimal Na$_{0.25}$HfNCl with $x_0 = 0.15$ from [13] are listed in Table 1. Reduced $T_C$ for $x > x_{opt}$ may also be influenced by structural disorder in Na$_x$HfNCl, owing to propensities for phase separation and intercalation staging [2]. Structural data show that $d_2$ is sharply reduced for the two non-optimal samples (i.e., $d_2 = 6.58$, 6.263, and 6.28 Å for $x = 0.25$, 0.28, and 0.29, respectively [13]), generally tending closer to the value 6.21 Å of the HfNCl host. The datum for Na$_{0.25}$HfNCl is shown in Fig. 1, lying just below the $T_{C0}$ line.

For the ammonia (NH$_3$) co-intercalated compounds, Li$_{0.2}$(NH$_3$)$_y$HfNCl with $T_C = 22.5$ K, Ca$_{0.11}$(NH$_3$)$_y$HfNCl with $T_C = 23$ K, and Eu$_{0.08}$(NH$_3$)$_y$HfNCl with $T_C = 23.6$ K from Meissner effect field-cooling data [8], the doping levels expressed as $vx = 0.2$, 0.22 and 0.24, for Li$^{+1}$, Ca$^{+2}$, and Eu$^{+3}$, respectively, are assumed optimal for each compound. However, measured values of $d_2$ for determining the ζ parameters are presently unavailable. For Ca$_x$(NH$_3$)$_y$HfNCl, the lattice parameter $c$ is found to be maximized at $x_{opt} = 0.11$ [9], indicating a maximized $d_2$ and behavior notably similar to the aforementioned case of Na$_x$HfNCl, where $d_2$ is maximized at 6.58 Å for $x = 0.25$ [2,3,13]. Hence this measured value of $d_2 = 6.58$ Å is assumed for these compounds for the purpose of estimating corresponding values of ζ from Eq. (1). Given this approximation, the calculated $T_{C0}$ values for the three NH$_3$ co-intercalated compounds are 21.44, 22.68, and 24.30 K, respectively. These results, included in Table 1, are plotted in Fig. 1 where four data points (Na$_{0.25}$HfNCl and $A_x$(NH$_3$)$_y$HfNCl) form a nearly linear cluster in close proximity to the $T_{C0}$ line of Eq. (2); the small deviations from the line suggest systematic corrections not well captured by assuming a non-varying $d_2$ for the three co-intercalates. It is also instructive to compare $Eu_{0.08}(NH_3)_yHfNCl$ and $Eu_{0.16}(NH_3)_yZrNCl$ [8], where the suppressed $T_C$ (=12.7 K) of the latter with factor of 2 larger x suggests severe overdoping.

Somewhat higher transition temperatures are observed for co-intercalation of Ca and Li with THF molecules. The compound $Ca_{0.11}(THF)_yHfNCl$ with $T_C = 26.0$ K and $d = 15.0$ Å is apparently close to being optimally doped ($x_{opt} \approx 0.11$), given the small systematic decreases in $T_C$ and $d$ for higher Ca contents [9]. For $Li_{0.2}(THF)_yHfNCl$ with $y \approx 0.2$ and $T_C = 25.6$ K (read from Fig. 3c in [20]), the slightly decreased $T_C$ reported for higher Li and THF contents indicate $x_{opt} \approx 0.2$ [20,35], the same as determined above for Li co-intercalated with $NH_3$. Depending upon the amount of co-intercalant y, $Li_x(THF)_yHfNCl$ generally forms both monolayers and bilayers of THF molecules within the van der Waals gap; $Li_{0.37}(THF)_yHfNCl$ with $y \approx 0.2$ and $d = 13.6$ Å is a monolayer form [20], whereas $Li_{0.48}(THF)_yHfNCl$ with $y \approx 0.3 - 0.4$ and $d = 18.7$ Å contains predominantly THF bilayers [20]. Both compounds are overdoped compositions, given $x > x_{opt}$ in addition to $x > y$, and since $T_C \approx 25.5$ K is reported for each compound, the basal-plane spacing $d$ cannot be considered a controlling parameter for the superconductivity [9,15]. Moreover, since both forms have comparable measures of basal-plane area $A$ and thickness $d_2$, their sharing the same $T_C$ points to an interaction distance $\zeta$ in $Li_x(THF)_yHfNCl$ that is essentially independent of $d$, and varies only slightly with doping.

Temperature-dependent [7]Li-NMR data on $Li_{0.48}(THF)_yHfNCl$ reveal a +5 ppm paramagnetic chemical shift [14], indicating Li covalently bonded with the oxygen of the THF molecule. In such a case, the charge is not localized at the Li site and one would expect that the interaction distance is determined by the $c$-axis projection of the $(Li-THF)^{+1}/Cl^{-1}$ dipole length, rather than by the intercalant layer midplane formulated in Eq. (1). Note that this type of chemical bond does not form between Li and $NH_3$, and its apparent absence in $Li_x(THF)_yTiNCl/Br$ and $Li_x(THF)_yZrNCl$ is likely associated with the comparatively larger basal-plane dimensions, affecting the occupation site symmetries, potentials, and energetics in a manner incompatible with a covalent Li-THF bond. In view of $T_C \approx 26$ K for optimally doped compounds $Ca_{0.11}(THF)_yHfNCl$ and $Li_{0.2}(THF)_yHfNCl$, an effective value $\zeta = 2.37$ Å may be deduced (again, assuming $x_0 = 0$) from Eq. (6), yielding estimates for $T_{C0}$ of 26.18 and 24.97 K, respectively (see Table 1). Without corroborating experimental evidence, however, the possible formation of (Li/Ca-THF)-Cl bonds remains in the realm of speculation.

Of the intercalated β-HfNCl compounds evaluated, $Na_{0.25}HfNCl$, $Li_{0.2}HfNCl$, $Li_{0.2}(NH_3)_yHfNCl$, $Ca_{0.11}(NH_3)_yHfNCl$, and $Eu_{0.08}(NH_3)_yHfNCl$, are found to be optimal, possessing measured $T_C$ values equal to their respective calculated $T_{C0}$ values within experimental uncertainty.

### 3.3 α-$A_x(S)_yTiNCl/Br$

Prior work [21] considered the intercalated α-$A_x(S)_yTiNCl$ series of compounds in [10] in terms of the RCS pair-breaking model noted in Section 2.2; this work extends the analysis to include the α-$A_x(S)_yTiNBr$ compounds reported in [11]. Table 2 contains data for seven $A_x(S)_yTiNCl$ compounds from [10] (excluding $Li_{0.13}TiNCl$ as in [10,21,22]) and six $A_x(S)_yTiNBr$ from [11]. In [10], $T_C$ for $Na_{0.16}(PC)_yTiNCl$ is tabulated as 7.4 K and indicated as 6.3 K by two-point extrapolation of $H_{C2}(T)$ in Fig. 2b of [10]. Transition temperatures $T_C$ and $d$ values for $Na_{0.16}(BC)_yTiNCl$ (BC is butylene carbonate, $C_5H_8O_3$), $K_{0.17}TiNCl$, $Rb_{0.24}TiNCl$, $Li_{0.13}(THF)_yTiNCl$, and $Li_{0.13}TiNCl$ are read from Fig. 3 of [10]. Data for

$Rb_{0.22}TiNBr$, $K_{0.21}TiNBr$, $Na_{0.23}TiNBr$, $Na_{0.21}(THF)_yTiNBr$, $Na_{0.21}(PC)_yTiNBr$, and $Li_{0.37}(THF)_yTiNBr$ are from Fig. 9 and Table 1 of [11]. The values of $T_{C0}$ in Table 2 are calculated from Eq. (6) under the caveat that $x = x_{opt}$ is assumed for each compound under discussion, since x-dependence data are presently unavailable, with $x_0 = 0$ inferred from [13], and $v = 1$.

A distinguishing geometrical feature of $\alpha$-$A_x(S)_yTiNCl/Br$ is found in the thickness $d_2$ of the $[MNX]_2$ block structure, which is typically ~1.1 Å smaller than for the $\beta$ form (compare Tables 1 and 2) [13]. Owing to the absence of refinement data, the $d_2$ of $\alpha$-TiNCl and $\alpha$-TiNBr [13] are used to obtain $\zeta$ for the co-intercalated $A_x(S)_yTiNCl$ compounds of [10] and the $A_x(S)_yTiNBr$ compounds of [11], respectively. For example, $\zeta$ for $Na_{0.16}(THF)_yTiNCl$ is computed from Eq. (1) as (13.105 Å – 5.183 Å)/2 = 3.9610 Å. Materials free of pair breaking and possessing the optimal cation doping necessarily exhibit $T_C = T_{C0}$ to within uncertainties. Note that the pairing model outlined in [23] and elsewhere [37-39] generally defines $T_C$ by the resistive zero when available, but given the broadened transitions typically prevalent in these materials, the magnetic onset $T_C$ (preferably field-cooled) value is acceptable. In any case, the suppression of $T_C$ below $T_{C0}$ evident in the remaining compounds can then be attributed to pair-breaking phenomena and treated accordingly.

As can be seen from $T_C \approx T_{C0}$ in Table 2, only the two compounds with the largest $\zeta$, $Na_{0.16}(PC)_yTiNCl$ and $Na_{0.16}(BC)_yTiNCl$, can be considered optimal (recall that $H_{C2}$ data give $T_C \approx 6.3$ K for $Na_{0.16}(PC)_yTiNCl$); these two compounds are represented in the plot of compounds with near optimal $T_C$ in Fig. 1. The remaining compounds with $T_C < T_{C0}$ show progressively larger deviations from optimal with decreasing $\zeta$. Interpreted in terms of pair-breaking, these deviations determine finite values of the pair-breaking parameter $\alpha$ as solutions of Eq. (4) with the results listed in Table 2. The two compounds with $T_C \approx T_{C0}$ are taken to have $\alpha = 0$. One finds that the $A_xTiNX$ compounds without co-intercalation molecules exhibit the highest $\alpha$ values, 1.07 – 1.78 meV, as expected for minimum $\zeta$. The pair-breaking rate associated with $2\alpha$, which is less than 3.6 meV, is therefore a very small component of the total scattering rate contained in $\hbar\tau^{-1}$. For example, considering the damping factors $\hbar\tau^{-1} > 0.2$ eV found optically for $Li_xZrNCl$ [46], and noting that transport measurements indicate higher resistivities for α-TiNCl-based compounds (e.g., $\rho(T_{peak}) \approx 0.27\ \Omega\text{cm}$ for $Na_{0.16}TiNCl$) when compared to β-ZrNCl-based compounds (e.g., $\rho(T_{peak}) \approx 6.2\ m\Omega\text{cm}$ for $Li_{0.08}ZrNCl$), these results are consistent with having $2\alpha << \hbar\tau^{-1}$, as expected.

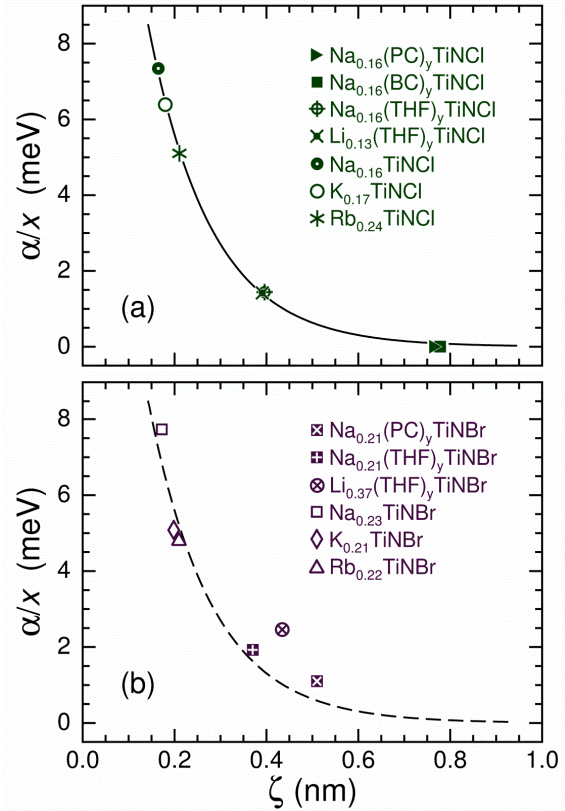

**Fig. 2** Pair breaking parameter $\alpha$ from Table 2 normalized to alkali content $x$ and plotted against interaction distance $\zeta$ for the $\alpha$-form compounds, **a** $A_x(S)_yTiNCl$ and **b** $A_x(S)_yTiNBr$. The *curves* represent the function of Eq. (5) fitted to the $A_x(S)_yTiNCl$ compounds.

In correspondence with the RCS pair-breaking model of Eq. (5), the pair-breaking parameters in Table 2 were scaled to doping and plotted as $\alpha/x$ vs. $\zeta$ in Fig. 2a, b for the $A_x(S)_y$TiNCl and $A_x(S)_y$TiNBr compounds, respectively. As shown in [21], the curve in Fig. 2a is a fit of the function $a_1\exp(-k_1\zeta)$ in Eq. (5) to the data for $A_x(S)_y$TiNCl with $a_1 = 23.9 \pm 1.0$ meV, $k_1 = 7.27 \pm 0.22\ nm^{-1}$, and fitting error of $\pm$ 0.10 meV. As $2\zeta$ approaches the van der Waals gap of pristine TiNCl (0.2618 nm) [13], $\alpha$ attains a hypothetical maximum of $(9.3 \pm 0.4\ \text{meV})x$, which is also small compared to reasonable estimates of $\hbar\tau^{-1}$. The attenuation factor can also be written as $k_1 \approx 0.84\ \pi A^{-1/2}$ with $\pi/A^{1/2} \approx 8.68\ nm^{-1}$ as averaged from Table 2.

While $\alpha/x$ determined for the $A_x(S)_y$TiNBr compounds also attenuates with $\zeta$, Fig. 2b reveals several notable differences with respect to $A_x(S)_y$TiNCl, which is represented by the dashed curve replicating the solid curve in panel a. The three data points for the $A_x$TiNBr compounds without co-intercalated molecules, which exhibit the strongest pair-breaking effect, indicate a 10 % level of consistency with the model for $A_x$TiNCl, although the attenuation with $\zeta$ appears slightly stronger. The $\alpha/x$ values for three co-intercalated $A_x(S)_y$TiNBr compounds display a non-monotonic trend with $\zeta$, mirroring a similar feature in the $T_C$ vs. $d^{-1}$ plot shown in Fig. 9 of [11]. These data also lie above the dashed curve in Fig. 2b, with $Li_{0.37}(THF)_y$TiNBr exhibiting the largest deviation. The Li doping for this compound substantially exceeds that of $Li_{0.13}(THF)_y$TiNCl, suggesting the presence of overdoping, i.e., x > $x_{opt}$. Calculated results for $T_{C0}$ and $\alpha$ would be smaller, relative to those presented in Table 2, for compounds where $x_{opt}$ is less than the actual doping x; e.g., $T_{C0}$ = 10.0 K and $\alpha$ = 0.261 meV are obtained for $Li_x(THF)_y$TiNBr if one assumes $x = x_{opt} = 0.13$. Note that the values $x_0 = 0$ and $x_{opt} = x$ have been provisionally used in constructing the entries for Table 2. In the case of the $A_x(S)_y$TiNBr compounds, assuming $x_0 > 0$ would reduce the calculated $T_{C0}$ and $\alpha/x$ values, perhaps resulting in behavior more commensurate with the exponential form of Eq. (5).

Noting the functional dependence of Eq. (1), the linear trend between $T_C$ and $d^{-1}$ in [10,22] is easily obtained by fitting the function $T_C = s/d$ to the $A_x(S)_y$TiNCl data (excluding $Li_{0.13}$TiNCl), yielding $s = 14.5 \pm 0.4$ nm K and rms deviation in $T_C$ of 0.87 K. Using this same function for the calculated $T_C$, as determined by Eq. (4) with $T_{C0}$ from Eq. (6) and $\alpha$ from Eq. (5), yields $s = 14.3 \pm 0.4$ nm K and rms deviation of 0.96 K. However, the rms deviation between the calculated $T_C$ and measured $T_C$ is only 0.54 K. For the 13 $A_x(S)_y$TiNCl/Br compounds shown in Fig. 9 of [11], the result for $s$ is statistically equivalent, although the rms deviation of $T_C$ from linearity in $d^{-1}$ increases to 1.25 K, while the rms deviation between calculated and measured $T_C$ is only 1.02 K. Thus the model for $T_C$ discussed herein provides significant improvement over scaling with $d^{-1}$.

Applicability of the pair-breaking model is illustrated by taking the experimental $T_C$ and the fitted function of Eq. (5) for $\alpha$, and then solving Eq. (4) for $T_{C0}$, yielding the hypothetical optimal transition temperature $T_{C0}^{(\alpha\rightarrow0)}$ for which the pair-breaking effect is removed. The results, given in Table 2, show that $T_{C0}^{(\alpha\rightarrow0)}$ for the five $A_x(S)_y$TiNCl compounds with $\alpha > 0$, have an rms deviation of 0.56 K from the calculated $T_{C0}$. Absent RCS-related pair-breaking, $T_C$ could approach 30 K for non-cointercalated $A_x$TiNCl compounds. Applying this analysis to the $A_x(S)_y$TiNBr compounds, exclusive of Li$_{0.37}$(THF)$_y$TiNBr and using the same fitted function for $\alpha$, yields results showing that $T_{C0}^{(\alpha \to 0)}$, given in Table 2, falls within 1 or 2 K of the corresponding optimal $T_{C0}$; assuming $x_{opt}$ = 0.13 for Li$_x$(THF)$_y$TiNBr gives $T_{C0}$ = 10.0 K and $T_{C0}^{(\alpha \to 0)}$ = 8.7 K.

Chlorine-deficient and alkali-intercalated $A_x$TiNCl$_{1-y}$ compounds have been prepared by heating $\alpha$-TiNCl in the presence of alkali-metal azides $A$N$_3$, resulting in compounds found to suffer from structural disorder and display lower $T_C$ relative to $A_x$TiNCl [10,12]. Similar processing of $\beta$-$M$NCl ($M$ = Zr, Hf) has been used to fabricate Cl-deintercalated polymorphs that also have slightly lower $T_C$, when compared to alkali intercalation and full Cl occupancy [12,55]. From the small changes in lattice spacings relative to the pristine $\beta$-forms, alkali intercalation is taken to be absent [12,55], although it is possible that residual $A_x$ forms an intercalation layer, where the cationic volume is compensated by the Cl$^{-1}$ vacancies.

## 4 Conclusions

The approach taken in the present work concerning $A_x(S)_y MNX$ provides a quantitative understanding of the observed superconductive behavior from the perspective of an interlayer Coulombic interaction model, with the relevant interaction distance $\zeta$ (different from $d$) measured between the cation $A_x$ and halogen $X$ layers of their respective mediating [$A_x(S)_y$] and superconducting [$MNX$]$_2$ charge reservoirs. From the optimal doping ($x_{opt} - x_0$) and considerations of charge allocation, the participating charge fraction $\sigma$ and optimal transition temperature $T_{C0}$ are then deduced. Since the length governing $T_C$ is $\zeta$ and not $d$, the anomalous behavior of Li$_x$TiNCl, H$_x$ZrNCl, and Li$_x$HfNBr is explained as being attributable to the location of the dopant such that a physically separated mediating layer, necessary for high-$T_C$ superconductivity, is not formed (i.e., $\zeta$ is unrealized).

Compared to the $\alpha$-TiN$X$-based compounds, the $\beta$-form materials boast higher $T_C$ values, differentiated primarily by sharper resistive transitions taken as evidence of more ordered intercalation structures. Both Li$_x$ZrNCl and Li$_x$HfNCl form solid solutions over a broad range of x, exhibiting a maximum in $T_C$ at x = $x_{opt}$, and an overdoped regime for x > $x_{opt}$ in which $T_C$ for Li$_x$HfNCl varies only slightly with x. This experimental result, coupled with changes in the $c$-axis variations associated with $x_{opt}$ and the continuous shift in phonon and plasma frequencies for x > $x_{opt}$, points to a transfer of charge primarily to the [$M$N]$_2$ layers. The excess of transferred charges is termed non-participating since they are benign with regard to the pairing interaction occurring between the Cl layer and Li intercalants; this explains why $T_C$ does not generally scale with x, $\omega_p$, or $\sigma_\mu$(T$\to$0). For compounds where the intercalated cation is covalently bonded with the THF co-intercalant, as may be the case for Li$_{0.2}$(THF)$_y$HfNCl and Ca$_{0.11}$(THF)$_y$HfNCl, it is speculated that $\zeta$ is not defined by the intercalant midplane, but instead determined by the bond between Cl$^{-1}$ anions and the charges distributed on (Li/Ca-THF) molecular cations.

The presence of strong scattering and broad superconducting transitions of the $\alpha$-TiN$X$ compounds suggest $T_C$ is suppressed by RCS pair breaking, postulated to arise from proximity of the superconducting layers to the disordered charges of the intercalation layer. Adapting a conventional pair-breaking model accordingly, the maximum attainable $T_C$ ($\leq T_{C0}$) for the $\alpha$-TiNCl-based compounds is shown to be well determined by a unique pair breaking function $\alpha$, which decreases exponentially with increasing $\zeta$. As one would expect, optimal superconductivity with $T_C \approx T_{C0}$ is indicated for the two $A_x(S)_y$TiNCl compounds with the largest interaction distances, $Na_{0.16}(PC)_yTiNCl$ and $Na_{0.16}(BC)_yTiNCl$ ($\zeta$ = 7.6738 and 7.7803 Å, respectively). For the other materials (apart from $Li_{0.13}TiNCl$) with smaller $\zeta$ parameters, RCS pair breaking is more dominant, yielding $T_C < T_{C0}$. The $\alpha$ function for $A_x(S)_yTiNBr$ is found to behave similarly, albeit with evidence for overdoping, x > $x_{opt}$, in $(S)_y$-containing compounds.

In all, nine compounds are found to be optimal, satisfying criteria of negligible pair-breaking and optimal doping: for $Na_{0.16}(PC)_yTiNCl$, $Na_{0.16}(BC)_yTiNCl$, $Li_{0.08}$-$ZrNCl$, $Li_{0.13}(DMF)_yZrNCl$, $Na_{0.25}HfNCl$, $Li_{0.2}HfNCl$, $Eu_{0.08}(NH_3)_yHfNCl$, $Ca_{0.11}(NH_3)_y$-$HfNCl$, and $Li_{0.2}(NH_3)_yHfNCl$, the calculated $T_{C0}$ agrees with the measured $T_C$ to within 0.8-K rms deviation. Combining these with previous results [23,37-39], the rms deviation between calculated $T_{C0}$ and measured $T_C$ is 1.35 K for 48 compounds from seven different high-$T_C$ families.

**Acknowledgments** The authors thank Dr. Shuai Zhang, Prof. Andrew M. Fogg, and Prof. Robert F. Marzke for providing helpful and important information. This work has been published [56].

**Compliance with Ethical Standards**

**Funding** This study was supported by Physikon Research Corporation (Project No. PL-206) and the New Jersey Institute of Technology.

**Conflict of Interest** The authors declare that they have no conflict of interest.